\documentclass[12pt, fullpage, a4paper]{article}
\usepackage{graphicx, latexsym}
\usepackage{setspace}
\usepackage{subfigure}
\usepackage{apacite}
\usepackage{amssymb, amsmath, amsthm}
\usepackage{bm}
\usepackage{epstopdf}
\usepackage{rotating} %used for \sidewaystable
\usepackage{apacite}
\usepackage{booktabs} %used for \toprule
\usepackage{multirow} %used for \multirow and \multicolumn
\usepackage{pifont}
\begin{document}
%\maketitle
\title{Joint distribution properties of Fully Conditional Specification under the normal linear model with normal inverse-gamma priors}
\author{Mingyang Cai, Stef van Buuren and Gerko Vink}
\date{}
\maketitle

\begin{abstract}
	Fully conditional specification (FCS) is a convenient and flexible multiple imputation approach. It specifies a sequence of simple regression models instead of a potential complex joint density for missing variables. However, FCS may not converge to a stationary distribution. Many authors have studied convergence properties of FCS when priors of conditional models are non-informative. We extend to the case of informative priors. This paper evaluates the convergence properties of the normal linear model with normal-inverse gamma prior. The theoretical and simulation results prove the convergence of FCS and show the equivalence of prior specification under the joint model and a set of conditional models when the analysis model is a linear regression with normal inverse-gamma priors.  
\end{abstract}

\section{Introduction}
Multiple imputation (Rubin, 1987) is a widely applied approach for the analysis of incomplete datasets. It involves replacing each missing cell with several plausible imputed values that are drawn from the corresponding posterior predictive distributions. There are two dominant approaches to arrive at those posterior distributions under multivariate missing data: joint modeling (JM) and fully conditional specification (FCS). 

Joint modeling requires a specified joint model for the complete data. Schafer\nocite{schafer1997analysis} (1997) illustrated joint modeling imputation under the multivariate normal model, the saturated multinomial model, the log-linear model, and the general location model. However, with an increasing number of variables and different levels of measurements, it can be challenging to formulate the joint distribution of the data. 

Fully conditional specification offers a solution to this challenge by allowing a flexible specification of the imputation model for each partially observed variable. The imputation procedure then starts by imputing missing values with a random draw from the marginal distribution. Each incomplete variable is then iteratively imputed with a specified univariate imputation model. 

Fully conditional specification has been proposed under a variety of names: chained equations stochastic relaxation, variable-by-variable imputation, switching regression, sequential regressions, ordered pseudo-Gibbs sampler, partially incompatible MCMC and iterated univariate imputation (van Buuren, 2018, Section 4.5.1)\nocite{van2018flexible}. Fully conditional specification can be of great value in practice because of its flexibility in model specification. FCS has become a standard in practice and has been widely implemented in software (e.g. \texttt{mice} and \texttt{mi} in \texttt{R}, \texttt{IVEWARE} in \texttt{SAS}, \texttt{ice} in \texttt{STATA} and module \texttt{MVA} in \texttt{SPSS}) \cite{buuren2010mice}.

Although many simulation studies demonstrated that fully conditional specification yields plausible imputations in various cases, the theoretical properties of fully conditional specification are not thoroughly understood \cite{van2007multiple}. A sequence of conditional models may not imply a joint distribution to which the algorithm converges. In such a case, the imputation results may systematically differ according to different visit sequences, which is named as ``order effects" \cite{hughes2014joint}. 

Van Buuren (2018, Section 4.6.1) stated two cases in which FCS converges to a joint distribution. First, if all imputation models are linear with a homogenous normal distributed response, the implicit joint model would be the multivariate normal distribution. Second, if three incomplete binary variables are imputed with a two-way interactions logistic regression model, FCS would be equivalent to the joint modeling under a zero three-way interaction log-linear model. Liu et al. (2013)\nocite{liu2014stationary} illustrated a series of sufficient conditions under which the imputation distribution for FCS converges in total variation to the posterior distribution of a joint Bayesian model when the sample size moves to infinity. Complementing the work of Liu et al., Hughes (2014) pointed out that, in addition to the compatibility, a ``non-informative margins" condition is another sufficient condition for the equivalency of FCS and joint modeling for finite samples. Hughes (2014) also showed that with multivariate normal distributed data and a non-informative prior, both compatibility and the non-informative margins conditions are satisfied. In that case, fully conditional specification and joint modeling provide imputations from the same posterior distribution. Zhu \& Raghunathan (2015) discussed conditions for convergence and assess properties of FCS. \nocite{zhu2015convergence}Many authors illustrated convergence properties of FCS when the prior for conditional models is non-informative. However, the case of informative priors has not received much attention. Therefore, we should consider the equivalent prior specification for informative priors under a sequence of conditional and corresponding joint models. This additional investigation allows the imputer to perform imputations under FCS even if they only collect the prior joint information for the incomplete dataset.      

For the initial step to evaluate convergence properties of FCS with informative priors, it is sensible to focus on the Bayesian normal linear models and the typical informative prior: normal inverse-gamma prior. This paper will briefly overview joint modeling, fully conditional specification, compatibility, and non-informative margins. Then, we derive a theoretical result and perform a simulation study to evaluate the non-informative margins condition. We also consider the prior for the target joint density of a sequence of normal linear models with normal inverse-gamma priors. Finally, some remarks are concluded.

\section{Background}
\subsection{Joint modeling}
Let $Y^{obs}$ and $Y^{mis}$ denote the observed and missing data in the dataset $Y$. Joint modeling involves specifying a parametric joint model $p(Y^{obs}, Y^{mis}|\theta)$ for the complete data and an appropriate prior distribution $p(\theta)$ for the parameter $\theta$. Incomplete cases are partitioned into groups according to various missing patterns and then imputed with different sub-models. Under the assumption of ignorability, the imputation model for each group is the corresponding conditional distribution derived from the assumed joint model 
\begin{equation*}
	p(Y^{mis}|Y^{obs}) = \int_{}p(Y^{mis}| Y^{obs}, \theta)p(\theta|Y^{obs})d\theta.
\end{equation*}
Since the joint modeling algorithm converges to the specified multivariate distribution, once the joint imputation model is correctly specified, results will be valid and theoretical properties are satisfactory. 

\subsection{Fully conditional specification}
Fully conditional specification attempts to define the joint distribution\\$p(Y^{obs}, Y^{mis}|\theta)$ by positing a univariate imputation model for each partially observed variable. The imputation model is typically a generalized linear model selected based on the nature of the missing variable (e.g. continuous, semi-continuous, categorical and count). Starting from some simple imputation methods, such as mean imputation or a random draw from the sampled values, FCS algorithms iteratively repeat imputations over all missing variables. Precisely, the \emph{t}th iteration for the incomplete variable \emph{$Y_{j}^{mis}$} consists of the following draws:
\begin{align*}
\theta_{j}^{t} \sim f(\theta_{j})f(Y_{j}^{obs}|Y_{-j}^{t-1}, \theta_{j})\\
Y_{j}^{mis(t)} \sim f(Y_{j}^{mis}|Y_{j}^{obs, }Y_{-j}^{t}, \theta_{j}^{t}),
\end{align*}
where $f(\theta_{j})$ is generally specified with a noninformative prior. After a sufficient number of iterations, typically ranging from 5 to 10 iterations (Van Buuren, 2018\nocite{van2018flexible}; Oberman et al., 2020\nocite{oberman2020missing}), the stationary distribution is achieved. The final iteration generates a single imputed dataset and the multiple imputations are created by applying FCS in parallel \emph{m} times with different seeds. If the underlying joint distribution defined by separate conditional models exists, the algorithm is equivalent to a Gibbs sampler. 

The attractive feature of fully conditional specification is the flexibility of model specification, which allows models to preserve features in the data, such as skip patterns, incorporating constraints and logical, and consistent bounds \cite{van2007multiple}. Such restrictions would be difficult to formulate when applying joint modeling. One could conveniently construct a sequence of conditional models and avoid the specification of a parametric multivariate distribution, which may not be appropriate for the data in practice.

\subsection{Compatibility} 
The definition of compatibility is given by Liu et al. (2014): let $Y = (Y_1, Y_2, \dots, Y_p)$ be a vector of random variables and $Y_{-j} = (Y_1, Y_2, \dots, Y_{j-1}, Y_{j+1}, \dots, Y_{p})$. A set of conditional models $\{f_{j}(Y_j|Y_{-j}, \theta_{j}) : \theta_{j} \in \Theta_{j}, j = 1, 2, \dots, p\}$ is said to be compatible if there exists a joint model $\{\emph{f}(Y|\theta) : \theta \in \Theta\}$ and a collection of surjective maps $\{t_{j} : \Theta \to \Theta_{j}\}$ such that for each $j$, $\theta_{j} \in \Theta_{j}$ and $\theta \in t_{j}^{-1}(\theta_{j}) = \{\theta : t_{j}(\theta) = \theta_{j}\}$. In that case 
\begin{gather*}
f_{j}(Y_j|Y_{-j}, \theta_{j}) = \emph{f}(Y_j|Y_{-j}, \theta).
\end{gather*}
Otherwise, $\{f_{j}, j = 1, 2, \dots, p\}$ is said to be incompatible.
A simple example of compatible models is a set of normal linear models for a vector of continous data: 
\begin{gather*}
Y_j = N((\textbf1, Y_{-j})\beta_{j}, \sigma_{j}^2), 
\end{gather*}
where $\beta_{j}$ is the vector of coefficients and $\textbf1$ is a vector of ones. In such a case, the joint model of $(Y_1, Y_2, \dots, Y_p)$ would be a multivariate normal distribution and the map $t_j$ is derived by conditional multivariate normal formula. On the other hand, the classical example for an incompatible model would be the linear model with squared terms (Liu et al., 2014; Barlett et al., 2015\nocite{bartlett2015multiple}).

Incompatibility is a theoretical weakness of fully conditional specification since, in some cases, it is unclear whether the algorithm indeed converges to the desired multivariate distribution (Arnold \& Press, 1989;\nocite{arnold1989compatible} Arnold et al., 2004;\nocite{arnold2004compatibility} Heckerman et al., 2000;\nocite{heckerman2000dependency} Van Buuren et al., 2006). Consideration of compatibility is significant when the multivariate density is of scientific interest. Both Hughes et al. (2014) and Liu et al. (2013) stated the necessity of model compatibility for the algorithm to converge to a joint distribution. Several papers introduced some cases in which FCS models are compatible with joint distributions (e.g., van Buuren, 2018; Raghunathan et al., 2001\nocite{raghunathan2001multivariate}). Van Buuren\nocite{van2006fully} (2006) also performed some simulation studies of fully conditional specification with strongly incompatible models and concluded the effects of incompatibility are negligible. However, further work is necessary to investigate the adverse effects of incompatibility in more general scenarios. 

\subsection{Non-informative margins}
Hughes et al. (2014) showed that the non-informative margins condition is sufficient for fully conditional specification to converge to a multivariate distribution. Suppose $\pi(\theta_{j})$ is the prior distribution of the conditional model $p(Y_j|Y_{-j}, \theta_{j})$ and $\pi(\theta_{-j})$ is the prior distribution of the marginal model $p(Y_{-j}|\theta_{-j})$, then the non-informative margins condition is satisfied if the joint prior could be factorized into independent priors $\pi(\theta_{j}, \theta_{-j}) = \pi(\theta_{j})\pi(\theta_{-j})$. It is worthwhile to note that the non-informative margin condition does not hold if $p(Y_j|Y_{-j}, \theta_{j})$ and $p(Y_{-j}|\theta_{-j})$ have the same parameter space. When the non-informative margins condition is violated, an order effect appears. In such a case, the inference of parameters would have systematic differences depending on the sequence of the variables in FCS algorithm. Simulations performed by Hughes et al. (2014) demonstrated that such an order effect is subtle. However, more research is needed to verify such claims, and it is necessary to be aware of the existence of the order effect. 

\section{Theoretical results}
This section proves the convergence of fully conditional specification under the normal linear model with normal inverse-gamma priors to a joint distribution. Since the compatibility of the normal linear model is well understood,  we will check the satisfaction of the non-informative margins condition. 

Starting with the problem of Bayesian inference for $\theta = (\mu, \Sigma)$ under a multivariate normal model, let us apply the following prior distribution. Suppose that, given $\Sigma$, the prior distribution of $\mu$ is assumed to be the conditionally multivariate normal,
\begin{equation}
\mu | \Sigma \sim N(\mu_{0}, \tau^{-1}\Sigma),
\end{equation}
where the hyperparameters $\mu_{0} \in \mathcal{R}^{p}$ and $\tau > 0$ are fixed and known and where $p$ denotes the number of variables. Moreover, suppose that the prior distribution of $\Sigma$ is an inverse-Wishart,
\begin{equation}
\Sigma \sim W^{-1}(m, \Lambda)
\end{equation}
for fixed hyperparameters $m \ge p$ and $\Lambda$. The prior density for $\theta$ can then be written as
\begin{equation}
\begin{array}{ll}
\pi(\theta) \propto &|\Sigma|^{-(\frac{m+p+2}{2})}\;\exp\;\{-\frac{1}{2}tr(\Lambda^{-1}\Sigma^{-1})\}\\
& \times\;\exp\;\{-\frac{\tau}{2}(\mu-\mu_{0})^{T}\Sigma^{-1}(\mu-\mu_{0})\}
\end{array}	
\end{equation}
For each variable $Y_{j}$, we partition the mean vector $\mu$ as $(\mu_j, \mu_{-j})^T$ and the covariance matrix $\Sigma$ as 
\begin{eqnarray*}
	\left(\begin{array}{cc}
		\omega_{j} & \xi_{j}^T\\
		\xi_{j} & \Sigma_{-j} 
	\end{array}\right),
\end{eqnarray*}
such that $Y_j \sim \mathcal{N}(\mu_j, \omega_{j})$ and $Y_{-j} \sim \mathcal{N}(\mu_{-j}, \Sigma_{-j})$. Similarly, we partition the scale parameter $\mu_{0}$ as $(\mu_{0j}, \mu_{0-j})^T$ and $\Lambda$ as
\begin{eqnarray*}
	\left(\begin{array}{cc}
		\Lambda_{j} & \psi_{j}^T\\
		\psi_{j} & \Lambda_{-j} 
	\end{array}\right).
\end{eqnarray*}
The conditional model of $Y_j$ given $Y_{-j}$ is the normal linear regression $Y_{j} = \alpha_j + \beta_{j}^TY_{-j} + \sigma_{j}$, where $\beta_{j}^T = \xi_{j}^T\Sigma_{-j}^{-1}$, $\alpha_j = \mu_j - \xi_{j}^T\Sigma_{-j}^{-1}\mu_{-j}$ and $\sigma_{j} = \omega_{j} - \xi_{j}^T\Sigma_{-j}^{-1}\xi_{j}$. The corresponding vectors of parameters $\theta_{j}$ and $\theta_{-j}$ would be 
\begin{equation}
\begin{array}{cc}
\theta_{j}  &= (\alpha_j, \beta_{j}, \sigma_{j})\\
\theta_{-j} &= (\mu_{-j}, \Sigma_{-j}).
\end{array}
\end{equation}
By applying the partition function illustrated by Eaton (2007, pp. 165)\nocite{10.2307/20461449} and by block diagonalization of a partitioned matrix, the joint prior for $\theta_{j}$ and $\theta_{-j}$ can be derived from $\pi(\theta)$ as :
\begin{equation}
\begin{array}{l}
\pi(\theta_{j}, \theta_{-j}) = p(\sigma_{j})p(\beta_{j}|\sigma_{j})p(\Sigma_{-j})\\
\times \exp\;\{-\frac{\tau}{2}(\alpha_{j} + \beta_{j}\mu_{0-j}\ - \mu_{0j})^{T}(\sigma_{j})^{-1}(\alpha_{j} + \beta_{j}\mu_{0-j}\ - \mu_{0j})\}\\
\times \exp\{-\frac{\tau}{2}(\mu_{-j}-\mu_{0-j})^{T}\Sigma_{-j}^{-1}(\mu_{-j}-\mu_{0-j})\} \times |\Sigma_{-j}|\\
=\pi(\theta_{j})\pi(\theta_{-j}),
\end{array}
\end{equation}
where
\begin{align}
&\pi(\theta_{j}) = p(\sigma_{j})p(\beta_{j}|\sigma_{j}) \nonumber\\
&\times \exp\;\{-\frac{\tau}{2}(\alpha_{j} + \beta_{j}\mu_{0-j}\ - \mu_{0j})^{T}(\sigma_{j})^{-1}(\alpha_{j} + \beta_{j}\mu_{0-j}\ - \mu_{0j})\},\\
&\pi(\theta_{-j}) = p(\Sigma_{-j})\times \exp\{-\frac{\tau}{2}(\mu_{-j}-\mu_{0-j})^{T}\Sigma_{-j}^{-1}(\mu_{-j}-\mu_{0-j})\} \times |\Sigma_{-j}|
\end{align}
and 

$p(\sigma_{j}) \sim W^{-1}(m, \lambda_j)$, $p(\beta_{j}|\sigma_{j}) \sim \mathcal{N}(\psi_{j}^T\Lambda_{-j}^{-1}, \lambda_j\Lambda_{-j}^{-1})$, $p(\Sigma_{-j}) \sim W^{-1}(m-1, \Lambda_{-j})$, $\lambda_j = \Lambda_{j} - \psi_{j}^T\Lambda_{-j}^{-1}\psi_{j}$ (Eaton, 2007, Section 8.2). Since the joint prior distribution factorizes into independent priors, the ``non-informative" margins condition is satisfied. Based on equations (6) and (7), we could derive the prior for the conditional linear model from the prior for the multivariate distribution:
\begin{equation}
\begin{array}{l}
p(\sigma_{j}) \sim W^{-1}(m, \lambda_j)\\
p(\beta_{j}|\sigma_{j}) \sim \mathcal{N}(\psi_{j}^T\Lambda_{-j}, \lambda_j\Lambda_{-j})\\
p(\alpha_{j}|\sigma_{j}) \sim \mathcal{N}(\mu_{0j} - \psi_{j}^T\Lambda_{-j}\mu_{02}, \tau^{-1}\sigma_{j} - (\mu_{-0j})^{2}\lambda_j\Lambda_{-j}^{-1})
\end{array}
\end{equation} 
Since the conditional $\beta_{j} | \sigma_{j}$ follows a normal distribution, the marginal distribution $\beta_{j}$ would be a student's t-distribution $\beta_{j} \sim t(\psi_{j}^T\Lambda_{-j}^{-1}, \\m\Lambda_{-j}^{-1}\lambda_{j}^{-1}, 2m-p+1)$. When the sample size increases, $\beta_{j}$ tends to the normal distribution $N(\psi_{j}^T\Lambda_{-j}^{-1}, \frac{\lambda_{j}\Lambda_{-j}}{m-1})$. Similarly, the marginal distribution $\alpha_{j}$ would be $t(\mu_{0j} - \psi_{j}^T\Lambda_{-j}\mu_{02}, m(\tau^{-1} - (\mu_{0-j})^{2}\Lambda_{-j}^{-1})\Lambda_{j}^{-1}, 2m-p+1)$. When the sample size increases, $\alpha_{j}$ tends to the normal distribution $N(\mu_{0j} - \psi_{j}^T\Lambda_{-j}\mu_{02},\\
 \frac{1}{(\tau^{-1} - (\mu_{-0j})^{2}\Lambda_{-j}^{-1})(m-1)}\Lambda_{j})$. Usually, when the sample size is over 30, the difference between student's t-distribution and the corresponding normally distributed approximation is negligble. With the prior transformation formula, one could apply Bayesian imputation under the normal linear model with normal inverse-gamma priors. This holds for both the prior information about the distribution of the data (e.g. location and scale of variables) and the scientific model (e.g. regression coefficients).  

\section{Simulation}
We perform a simulation study to demonstrate the validity and the convergence of fully conditional specification when the conditional models are simple linear regressions with an inverse gamma prior for the error term and a multivariate normal prior for regression weights. In addition, we look for the disappearance of order effects, which is evident in the convergence of fully conditional specification to a multivariate distribution. 

We repeat the simulation 500 times and generate a dataset with 200 cases for every simulation according to the following multivariate distribution :
\begin{eqnarray*}
	\begin{pmatrix}x\\
		y\\
		z
	\end{pmatrix} & \sim & \mathcal{N}\left[\left(\begin{array}{c}
		1\\
		4\\
		9
	\end{array}\right),\left(\begin{array}{ccc}
		4 & 2 & 2\\
		2 & 4 & 2\\
		2 & 2 & 9 
	\end{array}\right)\right]\\
\end{eqnarray*}
Fifty percent missingness is induced on either variable $x$, $y$ or $z$. The proportion of the three missing patterns is equal. When evaluating whether it is appropriate to specify a normal inverse gamma prior, we consider both missing completely at random (MCAR) mechanisms and right-tailed missing at random (MARr) mechanisms where higher values have a larger probability to be unobserved. When investigating the existence of order effects, we only conduct the simulation under MCAR missingness mechanism to ensure that the missingness does not attribute to any order effects. We specify a weak informative prior for two reasons. First, with a weak informative prior, the frequentist inference is still plausible by applying Rubin's rules (1987, pp.76). Second, Goodrich et al. (2019)\nocite{Goodrich2019} suggested that compared with flat non-informative priors, weak informative priors places warranted weight to extreme parameter values. In such a case, The prior under the joint model is specified as: $\mu_{0} = (0, 0, 0)^T$, $\tau = 1$, $m = 3$ and 
\begin{eqnarray*}
	\Lambda = \left(\begin{array}{ccc}
		60 & 0 & 0\\
		0 & 60 & 0\\
		0 & 0 & 60 
	\end{array}\right)
\end{eqnarray*}
and the corresponding prior for separated linear regression model would be the same, with $\pi(\sigma) \sim W^{-1}(3, 60)$ and 
\begin{eqnarray*}
	(\alpha, \beta)^T
	 & \sim & \mathcal{N}\left[\left(\begin{array}{c}
		0\\
		0\\
		0
	\end{array}\right),\left(\begin{array}{ccc}
		60 & 0 & 0\\
		0 & 3600 & 0\\
		0 & 0 & 3600 
	\end{array}\right)\right].\\
\end{eqnarray*}
\subsection{Scalar inference for the mean of variable Y}
The aim is to assess whether Bayesian imputation under a normal linear model with normal inverse gamma priors would yield unbiased estimates and exact coverage of the nominal 95\% confidence intervals. Table \ref{tab1} shows that with weak informative prior, fully conditional specification also provides valid imputations. The estimates are unbiased, and the coverage of the nominal 95\% confidence intervals is correct under both MCAR and MARr. Without the validity of a normal inverse gamma prior specification, further investigations into the convergence would be redundant. 
\begin{table}[h]
	\centering
	\vspace{0.5cm}
	\begin{tabular}{ccccc}
		& Bias  & Cov  & Ciw &  \\
		MCAR & 0     & 0.95 & 0.74 &  \\
		MARr & -0.01 & 0.97 & 0.73 &  \\
		&       &      &  & 
	\end{tabular}
\caption{Bias of the estimates ($E(Y)$) and coverage of nominal 95\% confidence intervals under MCAR and MARr}
\label{tab1}
\end{table}

\subsection{order effect evaluation}
\begin{figure}[h]
	\centering
	\includegraphics[scale=0.7]{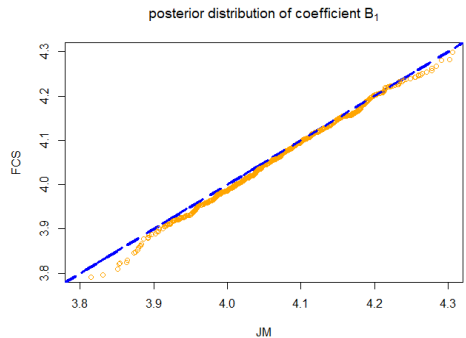}
	\caption{qqplot demonstrating the closeness of the posterior distribution of JM and FCS for $\beta_{1}$}
	\label{fig1}
\end{figure}

The visit sequence laid upon the simulation is $z$, $x$ and $y$. To identify the presence of any systematic order effect, we estimate the regression coefficient directly after updating variable $z$ and after updating variable $x$. Specifically, the \emph{i}th iteration of fully conditional specification would be augmented as:
\begin{enumerate}
	\item impute $z$ given $x^{i-1}$ and $y^{i-1}$.
	\item build the linear regression $y = \alpha + \beta_{1}x + \beta_{2}z + \epsilon$ and collect the coefficient $\beta_{1}$, donoted as $\hat{\beta_{1}}^z$.
	\item impute $x$ given $z^{i}$ and $y^{i-1}$.
    \item build the linear regression $y = \alpha + \beta_{1}x + \beta_{2}z + \epsilon$ and collect the coefficient $\beta_{1}$, donoted as $\hat{\beta_{1}}^x$.
    \item impute $y$ given $z^{i}$ and $x^{i}$.	 
\end{enumerate}
After a burn-in period with 10 iterations, the fully conditional specification algorithm was performed with an additional 1000 iterations, in which differences between the estimates $\hat{\beta_{1}}^z - \hat{\beta_{1}}^x$ are recorded. The estimates from the first 10 iterations are omitted since the FCS algorithms commonly reach convergence around 5 to 10 iterations. Estimates from the additional 1000 iterations would be partitioned into subsequences with equal size, which are used for variance calculation. We calculate the nominal 95\% confidence interval of the difference. The standard error of the difference is estimated with batch-means methods (Albert, 2009, pp.124)\nocite{albert2009bayesian}. The mean of $\hat{\beta_{1}}^z - \hat{\beta_{1}}^x$ is set to zero. Since only three 95\% confidence intervals derived from 500 repetitions do not cross the zero, there is no indication of any order effects. We also monitor the posterior distribution of the coefficient under both joint modeling and fully conditional specification. Figure \ref{fig1} shows a quantile-quantile plot demonstration the closeness of the posterior distribution for $\beta_{1}$ derived from both joint modeling and fully conditional specification. Since the posterior distributions for $\beta_{1}$ under joint modeling and FCS are very similar, any differences may be considered negligible in practice.  

All these results confirm that under the normal inverse gamma prior, Bayesian imputation under normal linear model converges to the corresponding multivariate normal distribution. 

\section{Conclusion}
Based on the theory of the non-informative margins condition proposed by Hughes et al. (2014), we prove the convergence of fully conditional specification under the normal linear model with normal-inverse-gamma prior distributions. Since it has been shown that a sequence of normal linear models is compatible with a multivariate normal density, we only focus on the non-informative margins condition for the prior. The transformation of the prior between a normal inverse gamma for fully conditional specification and a normal inverse Wishart for joint modeling is useful. With transformation, one could apply fully conditional specification and only collect prior information relative to the distribution of variables rather than scientific models.  

Fully conditional specification is an appealing imputation method because it allows one to specify a sequence of flexible and simple conditional models and bypass the difficulty of multivariate modeling in practice. The default prior for normal linear regression is Jeffreys prior, which satisfies the non-informative margin condition. However, it is worth developing other types of priors for fully conditional specification such that one could select the prior, which suits the description of prior knowledge best. Many researchers have discussed the convergence condition of FCS. However, there is no conclusion for the family of posterior distributions that satisfies the condition of convergence. In such a case, when including new kinds of priors in fully conditional specification algorithms, it is necessary to investigate the convergence of the algorithm with new posterior distributions. Specifically, one should study the non-informative margin conditions for new priors. Compatibility should also be considered if the imputation model is novel. Our work takes steps in this direction. 

Although a series of investigations have shown that the adverse effects of violating compatibility and the non-informative margin conditions may be small, all of these investigations rely on pre-defined simulation settings. More research is needed to verify conditions under which the fully conditional specification algorithm converges to a multivariate distribution and cases in which the violation of compatibility and non-informative margin has negligible adverse impacts on the result.

There are several directions for future research. From one direction, it is possible to develop a prior setting to eliminate order effects of the fully conditional specification algorithm under the general location model since the compatibility and non-informative margins conditions are satisfied under the saturated multinomial distribution. Moreover, various types of priors of the generalized linear model for the fully conditional specification and corresponding joint modeling rationales could be developed. Another open problem is the convergence condition and properties of block imputation, which partitions missing variables into several blocks and iteratively imputes blocks (Van Buuren, 2018, Sec. 4.7.2). Block imputation is a more flexible and user-friendly method. However, its properties have yet to be studied. Finally, it is necessary to investigate the implementation of prior specifications in software.   

\section{Data availability}
The datasets analysed during the current study available from the corresponding author on reasonable request and are available in the Github repository.

\newpage
\bibliographystyle{apacite}
\bibliography{pd}

\begin{thebibliography}{}

\bibitem [\protect \citeauthoryear {%
Albert%
}{%
Albert%
}{%
{\protect \APACyear {2009}}%
}]{%
albert2009bayesian}
\APACinsertmetastar {%
albert2009bayesian}%
\begin{APACrefauthors}%
Albert, J.%
\end{APACrefauthors}%
\unskip\
\newblock
\APACrefYear{2009}.
\newblock
\APACrefbtitle {Bayesian computation with R} {Bayesian computation with r}.
\newblock
\APACaddressPublisher{}{Springer}.
\PrintBackRefs{\CurrentBib}

\bibitem [\protect \citeauthoryear {%
Arnold%
, Castillo%
\BCBL {}\ \BBA {} Sarabia%
}{%
Arnold%
\ \protect \BOthers {.}}{%
{\protect \APACyear {2004}}%
}]{%
arnold2004compatibility}
\APACinsertmetastar {%
arnold2004compatibility}%
\begin{APACrefauthors}%
Arnold, B\BPBI C.%
, Castillo, E.%
\BCBL {}\ \BBA {} Sarabia, J\BPBI M.%
\end{APACrefauthors}%
\unskip\
\newblock
\APACrefYearMonthDay{2004}{}{}.
\newblock
{\BBOQ}\APACrefatitle {Compatibility of partial or complete conditional
  probability specifications} {Compatibility of partial or complete conditional
  probability specifications}.{\BBCQ}
\newblock
\APACjournalVolNumPages{Journal of statistical planning and
  inference}{123}{1}{133--159}.
\PrintBackRefs{\CurrentBib}

\bibitem [\protect \citeauthoryear {%
Arnold%
\ \BBA {} Press%
}{%
Arnold%
\ \BBA {} Press%
}{%
{\protect \APACyear {1989}}%
}]{%
arnold1989compatible}
\APACinsertmetastar {%
arnold1989compatible}%
\begin{APACrefauthors}%
Arnold, B\BPBI C.%
\BCBT {}\ \BBA {} Press, S\BPBI J.%
\end{APACrefauthors}%
\unskip\
\newblock
\APACrefYearMonthDay{1989}{}{}.
\newblock
{\BBOQ}\APACrefatitle {Compatible conditional distributions} {Compatible
  conditional distributions}.{\BBCQ}
\newblock
\APACjournalVolNumPages{Journal of the American Statistical
  Association}{84}{405}{152--156}.
\PrintBackRefs{\CurrentBib}

\bibitem [\protect \citeauthoryear {%
Bartlett%
, Seaman%
, White%
, Carpenter%
\BCBL {}\ \BBA {} Initiative*%
}{%
Bartlett%
\ \protect \BOthers {.}}{%
{\protect \APACyear {2015}}%
}]{%
bartlett2015multiple}
\APACinsertmetastar {%
bartlett2015multiple}%
\begin{APACrefauthors}%
Bartlett, J\BPBI W.%
, Seaman, S\BPBI R.%
, White, I\BPBI R.%
, Carpenter, J\BPBI R.%
\BCBL {}\ \BBA {} Initiative*, A\BPBI D\BPBI N.%
\end{APACrefauthors}%
\unskip\
\newblock
\APACrefYearMonthDay{2015}{}{}.
\newblock
{\BBOQ}\APACrefatitle {Multiple imputation of covariates by fully conditional
  specification: Accommodating the substantive model} {Multiple imputation of
  covariates by fully conditional specification: Accommodating the substantive
  model}.{\BBCQ}
\newblock
\APACjournalVolNumPages{Statistical methods in medical
  research}{24}{4}{462--487}.
\PrintBackRefs{\CurrentBib}

\bibitem [\protect \citeauthoryear {%
Buuren%
\ \BBA {} Groothuis-Oudshoorn%
}{%
Buuren%
\ \BBA {} Groothuis-Oudshoorn%
}{%
{\protect \APACyear {2010}}%
}]{%
buuren2010mice}
\APACinsertmetastar {%
buuren2010mice}%
\begin{APACrefauthors}%
Buuren, S\BPBI v.%
\BCBT {}\ \BBA {} Groothuis-Oudshoorn, K.%
\end{APACrefauthors}%
\unskip\
\newblock
\APACrefYearMonthDay{2010}{}{}.
\newblock
{\BBOQ}\APACrefatitle {mice: Multivariate imputation by chained equations in R}
  {mice: Multivariate imputation by chained equations in r}.{\BBCQ}
\newblock
\APACjournalVolNumPages{Journal of statistical software}{}{}{1--68}.
\PrintBackRefs{\CurrentBib}

\bibitem [\protect \citeauthoryear {%
Eaton%
}{%
Eaton%
}{%
{\protect \APACyear {2007}}%
}]{%
10.2307/20461449}
\APACinsertmetastar {%
10.2307/20461449}%
\begin{APACrefauthors}%
Eaton, M\BPBI L.%
\end{APACrefauthors}%
\unskip\
\newblock
\APACrefYearMonthDay{2007}{}{}.
\newblock
{\BBOQ}\APACrefatitle {Multivariate Statistics: A Vector Space Approach}
  {Multivariate statistics: A vector space approach}.{\BBCQ}
\newblock
\APACjournalVolNumPages{Lecture Notes-Monograph Series}{53}{}{i--512}.
\newblock
\begin{APACrefURL} \url{http://www.jstor.org/stable/20461449} \end{APACrefURL}
\PrintBackRefs{\CurrentBib}

\bibitem [\protect \citeauthoryear {%
Goodrich%
, Gabry%
, Ali%
\BCBL {}\ \BBA {} Brilleman%
}{%
Goodrich%
\ \protect \BOthers {.}}{%
{\protect \APACyear {2019}}%
}]{%
Goodrich2019}
\APACinsertmetastar {%
Goodrich2019}%
\begin{APACrefauthors}%
Goodrich, B.%
, Gabry, J.%
, Ali, I.%
\BCBL {}\ \BBA {} Brilleman, S.%
\end{APACrefauthors}%
\unskip\
\newblock
\APACrefYearMonthDay{2019}{}{}.
\newblock
\APACrefbtitle {rstanarm: {Bayesian} applied regression modeling via {Stan}.}
  {rstanarm: {Bayesian} applied regression modeling via {Stan}.}
\newblock
\begin{APACrefURL} \url{https://mc-stan.org/rstanarm} \end{APACrefURL}
\newblock
\APACrefnote{R package version 2.19.2}
\PrintBackRefs{\CurrentBib}

\bibitem [\protect \citeauthoryear {%
Heckerman%
, Chickering%
, Meek%
, Rounthwaite%
\BCBL {}\ \BBA {} Kadie%
}{%
Heckerman%
\ \protect \BOthers {.}}{%
{\protect \APACyear {2000}}%
}]{%
heckerman2000dependency}
\APACinsertmetastar {%
heckerman2000dependency}%
\begin{APACrefauthors}%
Heckerman, D.%
, Chickering, D\BPBI M.%
, Meek, C.%
, Rounthwaite, R.%
\BCBL {}\ \BBA {} Kadie, C.%
\end{APACrefauthors}%
\unskip\
\newblock
\APACrefYearMonthDay{2000}{}{}.
\newblock
{\BBOQ}\APACrefatitle {Dependency networks for inference, collaborative
  filtering, and data visualization} {Dependency networks for inference,
  collaborative filtering, and data visualization}.{\BBCQ}
\newblock
\APACjournalVolNumPages{Journal of Machine Learning Research}{1}{Oct}{49--75}.
\PrintBackRefs{\CurrentBib}

\bibitem [\protect \citeauthoryear {%
Hughes%
\ \protect \BOthers {.}}{%
Hughes%
\ \protect \BOthers {.}}{%
{\protect \APACyear {2014}}%
}]{%
hughes2014joint}
\APACinsertmetastar {%
hughes2014joint}%
\begin{APACrefauthors}%
Hughes, R\BPBI A.%
, White, I\BPBI R.%
, Seaman, S\BPBI R.%
, Carpenter, J\BPBI R.%
, Tilling, K.%
\BCBL {}\ \BBA {} Sterne, J\BPBI A.%
\end{APACrefauthors}%
\unskip\
\newblock
\APACrefYearMonthDay{2014}{}{}.
\newblock
{\BBOQ}\APACrefatitle {Joint modelling rationale for chained equations} {Joint
  modelling rationale for chained equations}.{\BBCQ}
\newblock
\APACjournalVolNumPages{BMC medical research methodology}{14}{1}{28}.
\PrintBackRefs{\CurrentBib}

\bibitem [\protect \citeauthoryear {%
Liu%
, Gelman%
, Hill%
, Su%
\BCBL {}\ \BBA {} Kropko%
}{%
Liu%
\ \protect \BOthers {.}}{%
{\protect \APACyear {2014}}%
}]{%
liu2014stationary}
\APACinsertmetastar {%
liu2014stationary}%
\begin{APACrefauthors}%
Liu, J.%
, Gelman, A.%
, Hill, J.%
, Su, Y\BHBI S.%
\BCBL {}\ \BBA {} Kropko, J.%
\end{APACrefauthors}%
\unskip\
\newblock
\APACrefYearMonthDay{2014}{}{}.
\newblock
{\BBOQ}\APACrefatitle {On the stationary distribution of iterative imputations}
  {On the stationary distribution of iterative imputations}.{\BBCQ}
\newblock
\APACjournalVolNumPages{Biometrika}{101}{1}{155--173}.
\PrintBackRefs{\CurrentBib}

\bibitem [\protect \citeauthoryear {%
Oberman%
, van Buuren%
\BCBL {}\ \BBA {} Vink%
}{%
Oberman%
\ \protect \BOthers {.}}{%
{\protect \APACyear {2020}}%
}]{%
oberman2020missing}
\APACinsertmetastar {%
oberman2020missing}%
\begin{APACrefauthors}%
Oberman, H\BPBI I.%
, van Buuren, S.%
\BCBL {}\ \BBA {} Vink, G.%
\end{APACrefauthors}%
\unskip\
\newblock
\APACrefYearMonthDay{2020}{}{}.
\newblock
{\BBOQ}\APACrefatitle {Missing the Point: Non-Convergence in Iterative
  Imputation Algorithms} {Missing the point: Non-convergence in iterative
  imputation algorithms}.{\BBCQ}
\newblock

\PrintBackRefs{\CurrentBib}

\bibitem [\protect \citeauthoryear {%
Raghunathan%
, Lepkowski%
, Van~Hoewyk%
, Solenberger%
\BCBL {}\ \protect \BOthers {.}}{%
Raghunathan%
\ \protect \BOthers {.}}{%
{\protect \APACyear {2001}}%
}]{%
raghunathan2001multivariate}
\APACinsertmetastar {%
raghunathan2001multivariate}%
\begin{APACrefauthors}%
Raghunathan, T\BPBI E.%
, Lepkowski, J\BPBI M.%
, Van~Hoewyk, J.%
, Solenberger, P.%
\BCBL {}\ \BOthersPeriod {.}\end{APACrefauthors}%
\unskip\
\newblock
\APACrefYearMonthDay{2001}{}{}.
\newblock
{\BBOQ}\APACrefatitle {A multivariate technique for multiply imputing missing
  values using a sequence of regression models} {A multivariate technique for
  multiply imputing missing values using a sequence of regression
  models}.{\BBCQ}
\newblock
\APACjournalVolNumPages{Survey methodology}{27}{1}{85--96}.
\PrintBackRefs{\CurrentBib}

\bibitem [\protect \citeauthoryear {%
Schafer%
}{%
Schafer%
}{%
{\protect \APACyear {1997}}%
}]{%
schafer1997analysis}
\APACinsertmetastar {%
schafer1997analysis}%
\begin{APACrefauthors}%
Schafer, J\BPBI L.%
\end{APACrefauthors}%
\unskip\
\newblock
\APACrefYear{1997}.
\newblock
\APACrefbtitle {Analysis of incomplete multivariate data} {Analysis of
  incomplete multivariate data}.
\newblock
\APACaddressPublisher{}{Chapman and Hall/CRC}.
\PrintBackRefs{\CurrentBib}

\bibitem [\protect \citeauthoryear {%
Van~Buuren%
}{%
Van~Buuren%
}{%
{\protect \APACyear {2007}}%
}]{%
van2007multiple}
\APACinsertmetastar {%
van2007multiple}%
\begin{APACrefauthors}%
Van~Buuren, S.%
\end{APACrefauthors}%
\unskip\
\newblock
\APACrefYearMonthDay{2007}{}{}.
\newblock
{\BBOQ}\APACrefatitle {Multiple imputation of discrete and continuous data by
  fully conditional specification} {Multiple imputation of discrete and
  continuous data by fully conditional specification}.{\BBCQ}
\newblock
\APACjournalVolNumPages{Statistical methods in medical
  research}{16}{3}{219--242}.
\PrintBackRefs{\CurrentBib}

\bibitem [\protect \citeauthoryear {%
Van~Buuren%
}{%
Van~Buuren%
}{%
{\protect \APACyear {2018}}%
}]{%
van2018flexible}
\APACinsertmetastar {%
van2018flexible}%
\begin{APACrefauthors}%
Van~Buuren, S.%
\end{APACrefauthors}%
\unskip\
\newblock
\APACrefYear{2018}.
\newblock
\APACrefbtitle {Flexible imputation of missing data} {Flexible imputation of
  missing data}.
\newblock
\APACaddressPublisher{}{CRC press}.
\PrintBackRefs{\CurrentBib}

\bibitem [\protect \citeauthoryear {%
Van~Buuren%
, Brand%
, Groothuis-Oudshoorn%
\BCBL {}\ \BBA {} Rubin%
}{%
Van~Buuren%
\ \protect \BOthers {.}}{%
{\protect \APACyear {2006}}%
}]{%
van2006fully}
\APACinsertmetastar {%
van2006fully}%
\begin{APACrefauthors}%
Van~Buuren, S.%
, Brand, J\BPBI P.%
, Groothuis-Oudshoorn, C\BPBI G.%
\BCBL {}\ \BBA {} Rubin, D\BPBI B.%
\end{APACrefauthors}%
\unskip\
\newblock
\APACrefYearMonthDay{2006}{}{}.
\newblock
{\BBOQ}\APACrefatitle {Fully conditional specification in multivariate
  imputation} {Fully conditional specification in multivariate
  imputation}.{\BBCQ}
\newblock
\APACjournalVolNumPages{Journal of statistical computation and
  simulation}{76}{12}{1049--1064}.
\PrintBackRefs{\CurrentBib}

\bibitem [\protect \citeauthoryear {%
Zhu%
\ \BBA {} Raghunathan%
}{%
Zhu%
\ \BBA {} Raghunathan%
}{%
{\protect \APACyear {2015}}%
}]{%
zhu2015convergence}
\APACinsertmetastar {%
zhu2015convergence}%
\begin{APACrefauthors}%
Zhu, J.%
\BCBT {}\ \BBA {} Raghunathan, T\BPBI E.%
\end{APACrefauthors}%
\unskip\
\newblock
\APACrefYearMonthDay{2015}{}{}.
\newblock
{\BBOQ}\APACrefatitle {Convergence properties of a sequential regression
  multiple imputation algorithm} {Convergence properties of a sequential
  regression multiple imputation algorithm}.{\BBCQ}
\newblock
\APACjournalVolNumPages{Journal of the American Statistical
  Association}{110}{511}{1112--1124}.
\PrintBackRefs{\CurrentBib}

\end{thebibliography}

\end{document}